# Anomalous connection between antiferromagnetic and superconducting phases in pressurized non-centrosymmetric heavy fermion compound CeRhGe$_3$

Honghong Wang[1,4]*, Jing Guo[1]*, Eric D. Bauer[2], Vladimir A Sidorov[3], Hengcan Zhao[1], Jiahao Zhang[1], Yazhou Zhou[1], Zhe Wang[1], Shu Cai[1], Ke Yang[4], Aiguo Li[4], Peijie Sun[1], Yi-feng Yang[1,5,6], Qi Wu[1], Tao Xiang[1,5,6], J. D. Thompson[2]†, Liling Sun[1,5,6]†

[1]*Institute of Physics, Chinese Academy of Sciences, Beijing 100190, China*
[2]*Los Alamos National Laboratory, MS K764, Los Alamos, NM 87545, USA*
[3]*Institute for High Pressure Physics, Russian Academy of Sciences, 142190 Troitsk, Moscow, Russia*
[4]*Shanghai Synchrotron Radiation Facilities, Shanghai Institute of Applied Physics, Chinese Academy of Sciences, Shanghai 201204, China*
[5]*University of Chinese Academy of Sciences, Beijing 100190, China*
[6]*Collaborative Innovation Center of Quantum Matter, Beijing, 100190, China*

Unconventional superconductivity frequently emerges as the transition temperature of a magnetic phase, typically antiferromagnetic (AFM), is suppressed continuously toward zero temperature. Here, we report contrary behavior in pressurized CeRhGe$_3$, a non-centrosymmetric heavy-fermion compound. We find that its pressure-tuned AFM transition temperature ($T_N$) appears to avoid a continuous decrease to zero temperature by terminating abruptly above a dome of pressure-induced superconductivity. Near 21.5 GPa, evidence for $T_N$ suddenly vanishes, the electrical resistance becomes linear in temperature and the superconducting transition reaches a maximum. In light of X-ray absorption spectroscopy measurements, these characteristics appear to be related to a pressured-induced Ce valence instability, which reveals as a sharp increase in the rate of change of Ce valence with applied pressure.

Evidence suggests that magnetism plays an important role for the emergence of unconventional superconductivity, with that superconductivity often develops in the vicinity of a sufficiently suppressed antiferromagnetically (AFM) ordered state [1-5], as demonstrated in the copper-oxide [6,7], iron-based [8,9] and heavy-electron superconductors [10,11]. A prominent common feature of their phase diagrams is that an AFM transition temperature ($T_N$) is continuously suppressed by pressure or chemical doping and presents a trend that it terminates at zero temperature, a magnetic quantum critical point, inside the superconducting phase. Over the past years, substantial efforts have been made to understand the interplay between AFM and superconducting phases, but it is still one of the most challenging issues in condensed matter physics.

Heavy-fermion materials provide a particular opportunity to study this issue because they are highly tunable with pressure, which does not introduce chemical/site disorder. Among heavy-fermion compounds, $CeTX_3$ (T = Co, Ir, Rh and X = Si, Ge) [12,13] possess an interesting crystal structure without inversion symmetry. In their pressure-induced superconducting state, these non-centrosymmetric compounds are expected to show unconventional pairing and corresponding exotic physics [14-18]. Indeed, superconductivity in $CeIrSi_3$, $CeRhSi_3$, $CeCoGe_3$ and $CeIrGe_3$ [19-23] develops near an antiferromagnetic boundary and displays unusual properties, including a very large upper critical field [19-25] and strong magnetic anisotropy [24,25]. Thus, the family of non-centrosymmetric superconductors provides a special platform to explore and understand the connection between the magnetic and superconducting states.

At ambient pressure, CeRhGe$_3$ is a heavy-electron antiferromagnet. Like all the other CeTX$_3$ (T = Co, Ir and X = Si, Ge) family members, it crystallizes in the tetragonal BaNiSn$_3$-type structure, space group I4/mm (No. 107) [13,18,26]. Our previous studies demonstrated that antiferromagnetic CeRhGe$_3$ becomes a superconductor at a pressure above 19 GPa and that substantial Kondo and spin-orbit coupling favor superconductivity in it as well as in the broader CeTX$_3$ family [27]. Here, we focus on an unusual relationship between the pressure dependence of $T_N$ and $T_C$ in CeRhGe$_3$ and the origin of a non-Fermi-liquid resistivity that appears once evidence for magnetic order vanishes.

Details of the crystal growth, characterization and high-pressure techniques can be found in Refs. [27,28]. Briefly, crystals of CeRhGe$_3$ were grown from a Rh$_{0.25}$Ge$_{0.75}$ eutectic self-flux. Their BaNiSn$_3$-type structure was confirmed by X-ray diffraction and studied as a function of pressure to 28.5 GPa on beamlines 15U at the Shanghai Synchrotron Radiation Facility and 4W2 at the Beijing Synchrotron Radiation Facility. A toroidal cell with a 3:2 mixture of glycerin and water as the pressure medium was used for pressure-dependent heat capacity measurements; whereas, magneto-electric and *ac* susceptibility measurements were made in a diamond anvil cell in which NaCl powder and silicon oil served as pressure medium, respectively. We discuss high pressure X-ray absorption measurements later.

Figure 1(a) displays the structure of CeRhGe$_3$, showing two distinct Ge sites in its unit cell that features the non-centrosymmetric structure. High-pressure heat capacity measurements, Fig. 1(b), reveal two anomalies that, like CeIrGe$_3$ [23], are associated with AFM transitions. We denote the higher AFM transition as $T_{N1}$ and the lower one by $T_{N2}$. The heat capacity jump at $T_{N1}$ is characteristic of a second order phase transition; whereas, a considerably smaller change in entropy accompanies the weaker anomaly at $T_{N2}$. Earlier high pressure resistance measurements up to 8 GPa confirmed $T_{N1}(P)$ [29], and signatures of both transitions clearly appeared in $\partial^2 R/\partial T^2$ for P < 13.7 GPa allow determining both $T_{N1}(P)$ and $T_{N2}(P)$ [27]. At higher pressure, only a single AFM transition is detectable, we thus label this transition as $T_N$. At 19.6 GPa, $R$(T) begins a pronounced drop at 1.3 K and that drop moves to lower temperatures with applied magnetic field [inset of Fig. 2(a)]. This field dependence of the resistivity drop and the appearance of a diamagnetic *ac* susceptibility in the second crystal at a slightly lower pressure [inset of Fig. 2(b)] indicate the presence of superconductivity.

These experiments lead to the pressure-temperature phase diagram in Fig. 3(a). Within the context of Doniach's model of competing Ruderman–Kittel–Kasuya–Yosida (RKKY) and Kondo interactions, the bell-shaped response of antiferromagnetism to pressure is consistent with a pressure-induced enhancement of the magnetic exchange $J$ [30] and, through the Shrieffer-Wolff transformation [31], to an associated increase in the square of the matrix element that mixes conduction and *f*-electron wave functions, $J \propto \langle V_{fc} \rangle^2$. A smooth extrapolation of $T_N(P)$ from pressures below ~17 GPa to higher pressures suggests that there might be a magnetic quantum critical point (QCP) near 20-22 GPa where it would be hidden by the dome of superconductivity. In contrast to this possibility and expectations from Doniach's model, $T_N(P)$ remains almost constant at higher pressures, being greater than $T_C(P)$ of the coexisting superconductivity, and resistive evidence for it disappears abruptly at $P_C \approx 21.5$ GPa where $T_C$ reaches a maximum. Such an anomalous connection between the AFM and superconducting phases in CeRhGe$_3$ is contrary to what is found in most strongly correlated electron superconductors whose Néel temperature is suppressed continuously into their superconducting dome [1-4,9,10]. Our results suggest that a magnetic quantum critical point is likely avoided. This relationship between magnetism and superconductivity, reproduced in a second crystal of CeRhGe$_3$, also is observed in a sister compound CeIrGe$_3$ [23]. For comparison, we plot the temperature-pressure phase diagram for CeIrGe$_3$ in Fig.3(a). Though $T_{N1}$ and $T_{N2}$ are lower in CeIrGe$_3$ than CeRhGe$_3$, they also merge at a critical pressure and superconductivity emerges at almost the same pressures as in CeRhGe$_3$.

Once evidence for magnetic order disappears, the resistance from $T_C$ to at least 10 K becomes $T$-linear, which is illustrated in Figs.3 (b) and (c). This is shown more clearly in Fig. 4(a) where we plot the temperature exponent $n$ derived from a logarithmic derivative, $\partial \ln(R(T)-R_0)/\partial \ln T$, of $R(T) = R_0 + AT^n$. The obviously non-quadratic temperature dependence, with $n \approx 1$, is typical of a non-Fermi-liquid that arises near a QCP [32] and that also is found in CeIrGe$_3$ above its critical pressure [23]. In the same pressure and temperature range, the non-Fermi-liquid resistance of CeRhGe$_3$ is comparably well fit to $R(T) = R_0 + AT + BT^2$, as it is in the cuprates [33]. (The fitted curves are indistinguishable from those shown in Figs. 3(b) and (c)).

Fitting parameters $R_0$, $A$ and $B$ are plotted in Figs. 4 (b-d). $R_0$ and $A$ tend to diverge upon approaching $P_C$ where evidence for $T_N$ disappears and $T_C$ reaches a maximum, but the quadratic coefficient $B$ tends to decrease at higher pressures. If superconductivity were suppressed and the resistance measured to lower temperatures, it is possible that the trend in $B$(P) might be reversed, but this remains to be determined. Taking the data in Figs. 4(b-d), however, it implies two coexisting scattering channels at $P \geq P_C$ [33] and that the source of scattering in the $T$-linear channel favors superconductivity. In CeIrGe$_3$, its linear-in-temperature resistance and pressure-induced superconductivity are observed [23], for which there is no direct evidence to indicate the superconductivity is driven by a magnetic QCP.

Besides magnetic criticality, critical valence fluctuations also are predicted to induce unconventional superconductivity [34] and a $T$-linear resistivity [35]. Further, when $T_N$ is reduced toward $T = 0$, critical fluctuations of the valence can terminate antiferromagnetic order in a first order transition [36]. In light of these predictions, their possible applicability to account for observations in Fig. 3 and the expected increase in $\langle V_{fc}\rangle^2$ at high pressures, we performed room-temperature L$_{III}$-edge X-ray absorption spectroscopy measurements, which are sensitive to changes in the 4$f$ configuration of CeRhGe$_3$ as a function of pressure. Experiments were carried out in a diamond anvil cell with low birefringent diamonds at beamline 15U of the Shanghai synchrotron radiation facility [37]. Results of this work are presented in Fig.5(a) in which data have been corrected for background and fluorescence contributions at each pressure. The relative intensity of each curve is normalized to zero at 5700 eV which is well below the $4f^1$ signal. At the lowest pressure, 3.55 GPa, there is no spectral weight within experimental resolution in the $4f^0$ channel, but increasing pressure induces intensity at the $4f^0$ energy and a decrease in $4f^1$ intensity. From these data, we calculate the mean valence (*v*) as a function of pressure, which is plotted in Fig. 5(b). As seen in Fig. 5(a) and (b), there is a spectral weight transfer from $4f^1$ to $4f^0$ under pressure, and the slope of pressure dependent mean valence changes abruptly at $P_C$ where the superconductivity develops. These results imply that CeRhG$_3$ has a remarkable valence instability at $P_C$, separating a magnetically ordered state and a mixed-valence state. These results indicate that the valence instability may play a non-trivial role in inducing superconductivity and non-Fermi-liquid properties near $P_C$ where evidence for magnetic order disappears at $T > T_C(P)$.

In summary, we have investigated the effect of pressure on the AFM and superconducting transitions of CeRhGe$_3$ through high-pressure heat capacity, resistance, *ac* susceptibility and L$_{III}$-edge absorption measurements. From these experiments we find an unusual relationship between antiferromagnetic and superconducting states in which a magnetic QCP is likely avoided. This relationship is contrary to that found in many strongly correlated systems, including the CeTX$_3$ family (T = Co, Ir and X = Si, Ge) with the exceptions of strongly mixed-valence CeCoSi$_3$ (which is not a superconductor) and CeIrGe$_3$ whose *T-P* phase diagram is similar to that of CeRhGe$_3$ [38]. Our observations, in addition to the largest residual resistance and lack of divergence in the $T^2$ coefficient of resistance at $P_C$, are consistent with a scenario in which a first-order valence transition or crossover at low temperatures produces a non-Fermi liquid resistance and abrupt loss of a signature for magnetic order at $P_C$ [36]. Indeed, the varied relationships among magnetism, criticality and superconductivity that are found in CeTX$_3$ are anticipated theoretically in this model of critical valence fluctuations and their interplay with magnetic order in heavy-fermion metals.

**Acknowledgements**

We thank Prof. Frank Steglich for fruitful discussions. Work in China was supported by the National Key Research and Development Program of China (Grant No. 2017YFA0302900, 2017YFA0303103, 2016YFA0300300 and 2015CB921303), the NSF of China (Grants No. 91321207, No. 11427805, No. 11404384, No. U1532267, No. 11604376, No. 11522435, No. 11774401), the Strategic Priority Research Program (B) of the Chinese Academy of Sciences (Grant No. XDB07020300). Work at Los Alamos National Laboratory was performed under the auspices of the U.S. DOE, Office of Basic Energy Sciences, Division of Materials Sciences and Engineering.

† Correspondence and requests for materials should be addressed to L.S. (llsun@iphy.ac.cn) and J.T (jdt@lanl.gov).

* These authors contributed equally to this work.

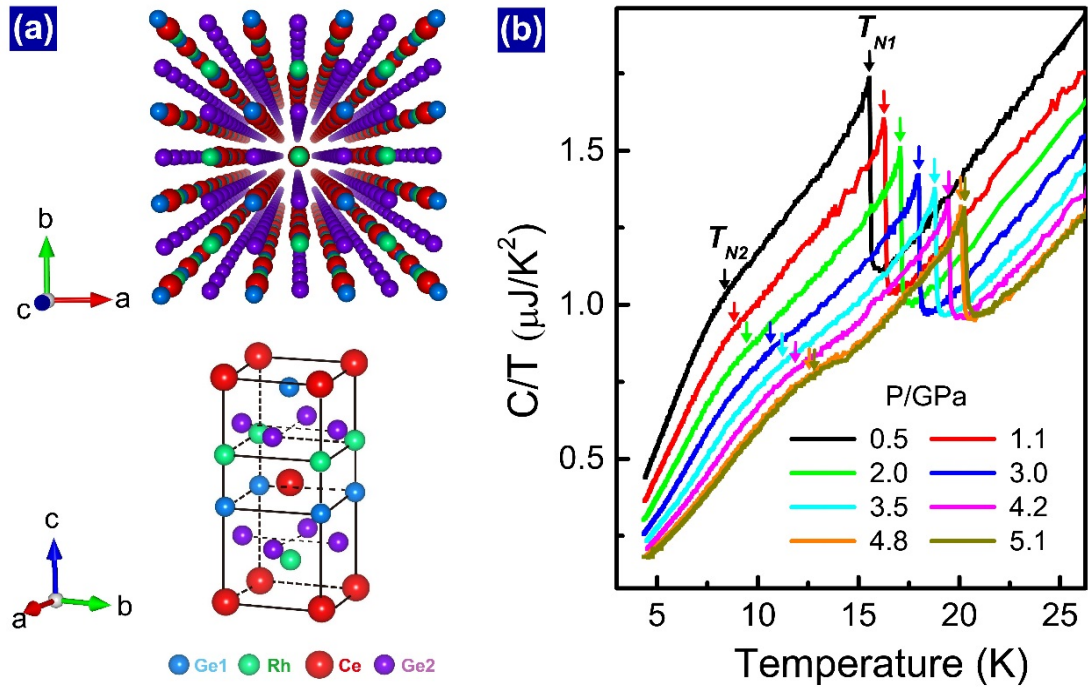

**Figure 1 Structure and high-pressure heat capacity results of CeRhGe$_3$.** (a) Illustration of crystal structure. The upper panel displays a view of the crystal structure along the tetragonal *c*-axis and lower panel shows a 3-dimensional representation of that structure. (b) Temperature dependence of heat capacity divided by temperature (*C/T*) at different pressures. $T_{N1}$ and $T_{N2}$ correspond to the higher and lower AFM transition temperatures, respectively.

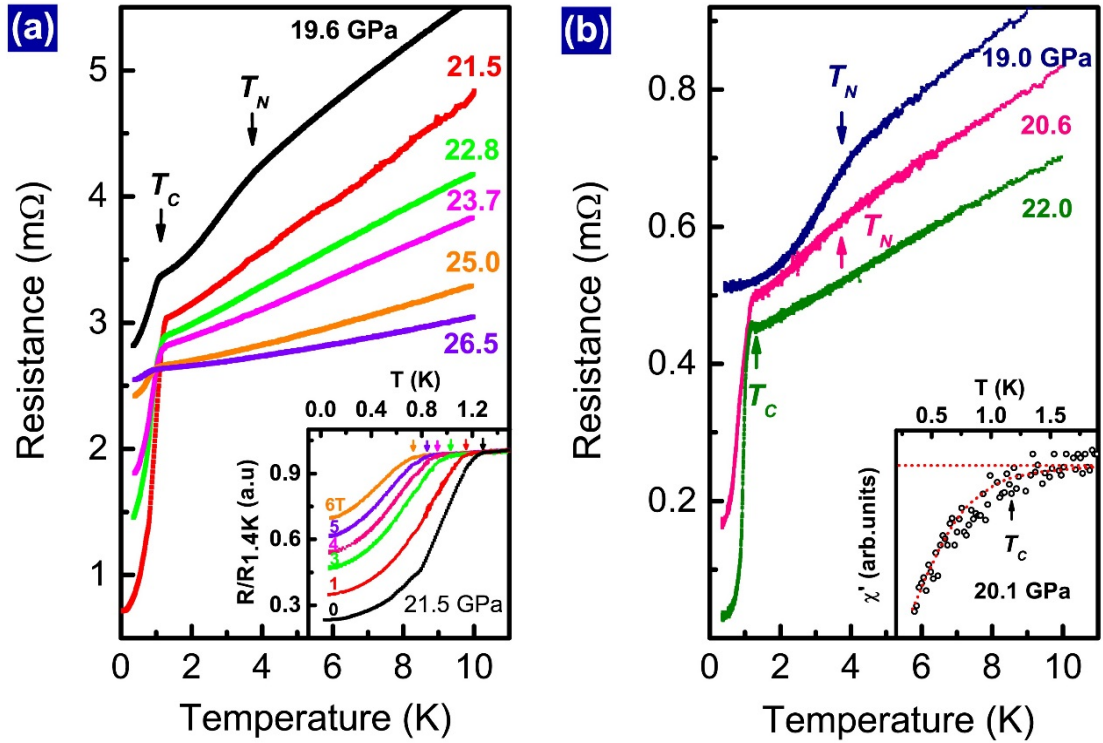

**Figure 2 Temperature dependent resistance of CeRhGe$_3$ at representative pressures.** Results are essentially identical for measurements on two crystals. The lack of zero resistance below the superconducting transition at $T_C$ is due to pressure-induced micro-cracks that are somewhat different in each crystal. $T_N$ marks the single Néel temperature that is found above ~13.7 GPa. The insets of (a) and (b) display the field dependence of $T_C$ at 21.5 GPa and a diamagnetic response starting at 1.2 K in *ac* susceptibility measurements at 20.1 GPa, respectively.

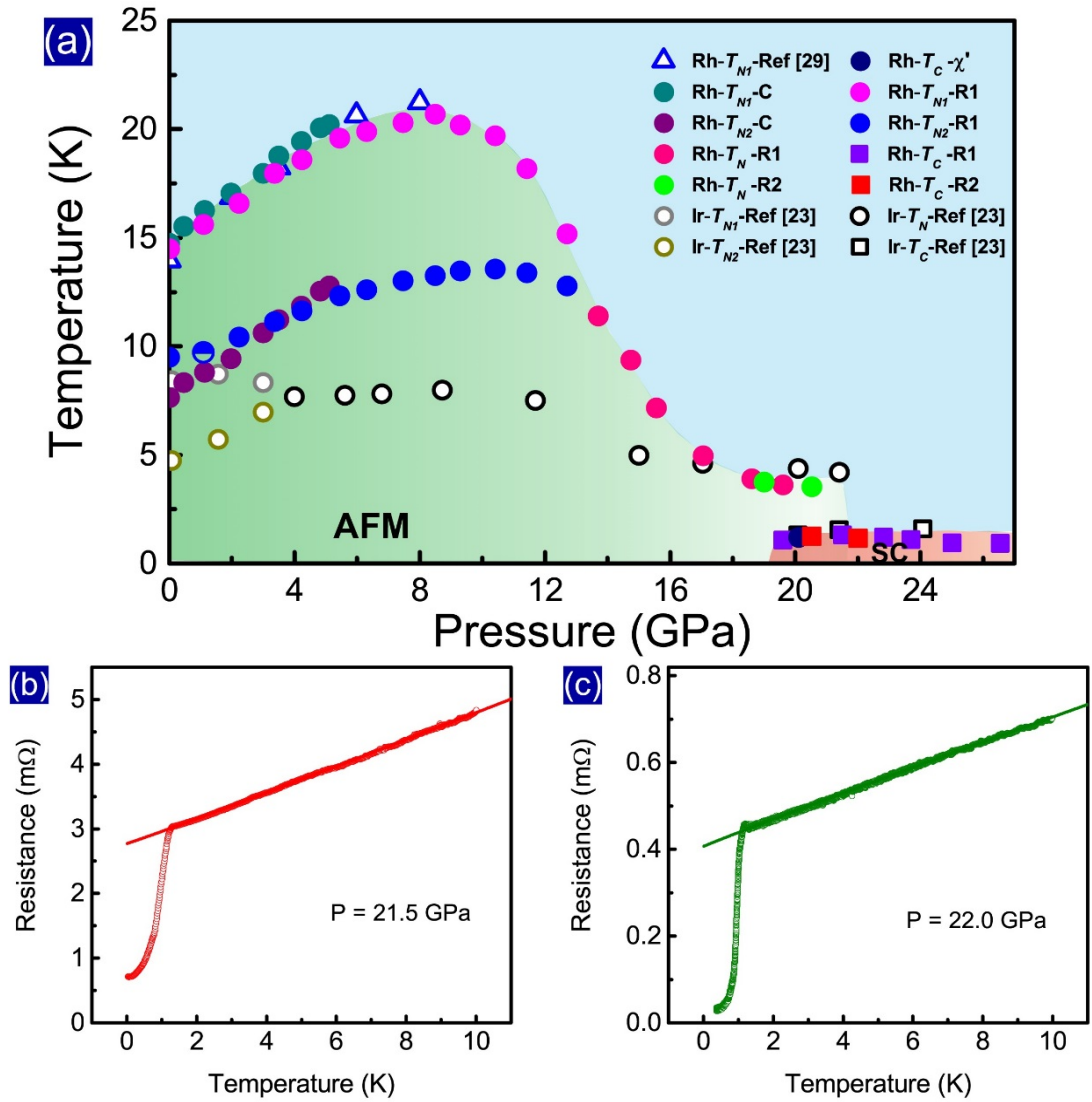

**Figure 3 Temperature-pressure phase diagram and resistance versus temperature for CeRhGe$_3$.** (a) Evolution of the AFM transition temperatures $T_{N1}$, $T_{N2}$ and $T_N$ and superconducting transition temperature $T_C$ with pressure for CeRhGe$_3$ and CeIrGe$_3$. The solid circles and squares are data obtained from GeRhGe$_3$ in this study, while the open triangles are from measurements on GeRhGe$_3$ reported in Ref. [29]. The open circles and squares correspond to transitions in CeIrGe$_3$ taken from Ref. [23]. Panels (b) and (c) show that the resistance above $T_C$ increases approximately linearly with temperature at pressures near and above $P_C \approx 21.5$ GPa.

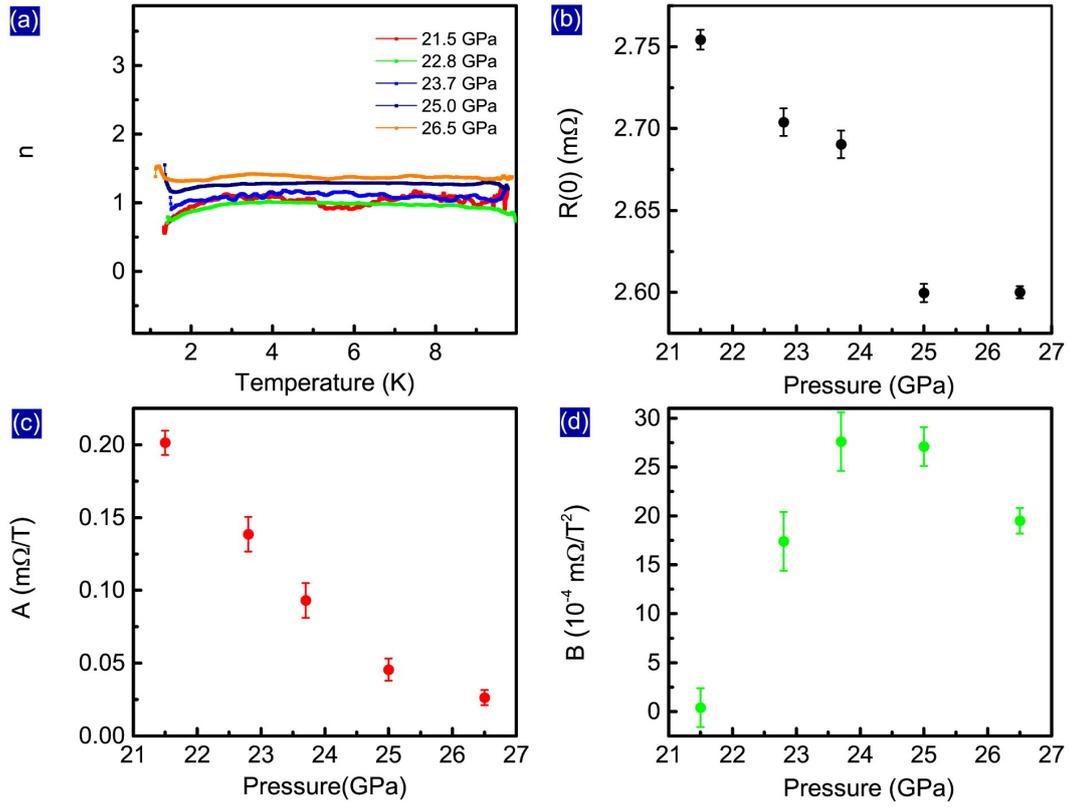

**Figure 4 Parameters characterizing the low temperature resistance of CeRhGe$_3$ at *P* > *P$_C$*.** (a) Exponent *n* of a power-law temperature variation of the resistance determined from a logarithmic derivative $\partial \ln(R(T)-R_0)/\partial \ln T$, assuming $R(T) = R_0 + AT^n$. (b)-(d) Parameters obtained from fitting the resistance to $R(T) = R_0 + AT + BT^2$. See text for details.

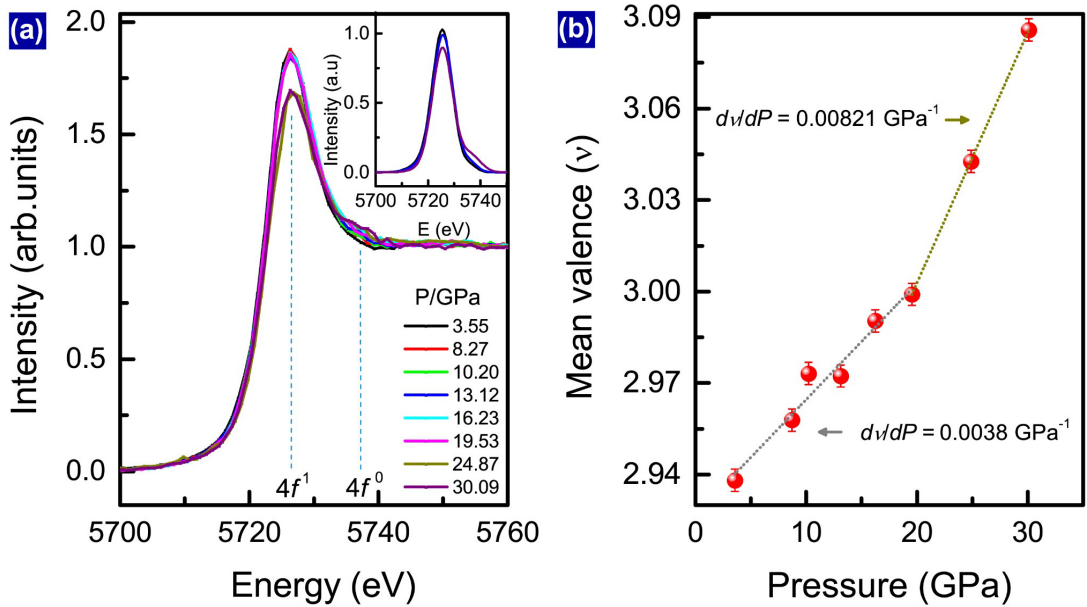

**Figure 5 Results of high-pressure X-ray absorption measurements.** (a) Ce-$L_{III}$ X-ray absorption spectra of CeRhGe$_3$ at various pressures and room temperature. The obvious drop in intensity at the *4f$^1$* line appears between 19.53 and 24.87 GPa. Pressure in the diamond cell was determined by the standard ruby-fluorescence technique. (b) Pressure dependence of the mean valence determined, as discussed in the text, from data in (a). Above $P_C$, the rate of change in valence is over twice that below $P_C$.